\documentstyle[fullpage,graphics,twocolumn]{article}
\title{The Hopgrid algorithm: multilevel synthesis of multigrid and wavelet theory}

\author{D. Ye\c{s}illeten and T. A. Arias\\
Department of Physics\\
Massachusetts Institute of Technology\\
Cambridge, Massachusetts 02139}

\date{August 8, 1996}

\begin{document}

\maketitle

\begin{abstract}
The multigrid algorithm is a multilevel approach to accelerate the
numerical solution of discretized differential equations in physical
problems involving long-range interactions.  Multiresolution analysis
of wavelet theory provides an efficient representation of functions
which exhibit localized bursts of short length-scale behavior.
Applications such as computing the electrostatic field in and around a
molecule should benefit from both approaches.  In this work, we
demonstrate how a novel interpolating wavelet transform, which in
itself is the synthesis of finite element analysis and wavelet theory,
may be used as the mathematical bridge to connect the two approaches.
The result is a specialized multigrid algorithm which may be applied
to problems expressed in wavelet bases.  With this approach,
interpolation and restriction operators and grids for the multigrid
algorithm are predetermined by an interpolating multiresolution
analysis.  We will present the new method and contrast its efficiency
with standard wavelet and multigrid approaches.
\end{abstract}

\section{Introduction} \label{intro}

Many applications such as electronic structure calculations require an
efficient method to solve discretized linear differential equations
which are generators of long range interactions, such as Poisson's
equation.  In modern electronic structure calculations, except
those using planewaves, solving Poisson's equation takes up a
significant amount of computational time.   It is essential to produce
an algorithm that has the capacity to solve efficiently problems
involving such long range interactions.

The multigrid algorithm is the state of the art in solving discretized
linear differential equations describing long range interactions.  It
uses a basic iterative method, such as the weighted Jacobi method,
over a sequence of scales to iterate faster toward the exact solution.
In some cases, even multigrid does not produce sufficiently rapid
convergence, particularly in calculations using wavelet bases.

Multiresolution analysis, from wavelet theory, provides the ability to
carry out calculations on non-uniform grids, focusing resolution only
in regions where it is needed.   It is especially useful if the local
region which requires high resolution moves through space during the
calculation, as when the atoms move in a molecule in an electronic
structure calculation.   However, a basis set alone does not provide
computational efficiency.

Often, a finite element basis set provides the analytic framework for
multigrid calculations.  Combining the advantages of finite element
analysis and the multigrid algorithm with wavelet theory, could
provide a very powerful method to solve discretized linear
differential equations.  This present research aims to unify these
approaches for use in a wide range of applications, especially in
electronic structure calculations, where all of the aforementioned
approaches prove to be very beneficial in various stages of the
calculations individually.

A basis set that combines wavelet theory with finite element analysis
exists and is known as the interpolet basis (Interpolets is based upon
Deslauriers and Dubuc scaling functions [4], which play the role of
both scaling functions and interpolets in the interpolet basis. See
also [5]).  The present research forms a natural bridge in the
synthesis of the multigrid algorithm with this basis, thereby adding
the power of multiresolution analysis to the multigrid algorithm.  In
particular, we introduce this new {\em Hopgrid} algorithm for use in
problems where an underlying wavelet or interpolet basis is
necessitated by the problem at hand and application of multigrid
algorithm could provide very useful in solving linear differential
equations in the problem.

In Section \ref{mgrid} of the paper, to establish a common notation,
we give a brief overview of the traditional multigrid algorithm and
the iterative methods that form the framework for this algorithm.
Section \ref{interp} introduces the interpolet theory.  Section
\ref{mra} includes the description of non-orthogonal multiresolution
analysis from wavelet theory in the interpolet basis.  Section
\ref{img} presents the basic ideas behind expressing the multigrid
algorithm in the interpolet basis, and describes how such a union
becomes more efficient by using multiresolution analysis.  Section
\ref{hgrid} introduces the new algorithm formed by a more impact
synthesis of the multigrid algorithm and the interpolet basis, the
Hopgrid Algorithm.  The last section expresses the results obtained
using this new algorithm.  The appendix of this paper aims to
familiarize the reader with the essentials of the multigrid algorithm
with a thorough discussion of the basic iterative methods and the
framework of this algorithm.

\section{Multigrid Algorithm} \label{mgrid}

This section briefly summarizes the multigrid algorithm [2] and the
interpolation and restriction operators used in this algorithm.  A
more detailed description of the multigrid algorithm and these
operators, aimed for the physicist, may be found in Section
\ref{appenmg}.  Throughout this section, superscripts are used to
refer to the scale on which a particular level of the problem is
solved.

\subsection{Full V-cycle Multigrid Algorithm}

The multigrid algorithm is used to solve discretized linear
differential equations of the form,
\begin{eqnarray} \label{m1}
Au = f,
\end{eqnarray}
where $A$ is the linear operator, $u$ is the solution vector, and $f$
is the source vector.  Multigrid is based upon a basic iterative
method, which is then applied over a sequence of scales in order to
improve the convergence rate.  One such iterative method used for this
purpose is the {\em weighted Jacobi method}, which provides the
following recursion,
\begin{eqnarray} \label{m2}
v_{(n+1)} = v_{(n)} + wD^{-1}r_{(n)}.
\end{eqnarray}
Here, $w$ is an appropriate weight (usually chosen to be $w=2/3$),
$v_{(n)}$ is the initial guess, $D$ is the diagonal matrix composed of
the diagonal elements of $A$, and $r_{(n)}$ is the {\em residual} at
step $n$, defined as $r_{(n)} = f - A v_{(n)}$.  We refer to
application of this recursion as {\em relaxation}.

Multigrid exploits this basic iterative method in a multiscale fashion
through the following procedure.  First, $\alpha$ relaxations are
performed on the scale on which the problem is defined.  Then, the
residual $r^{(0)}$ is transfered up to a spatially coarser scale,
using a linear {\em restriction} operator, $r^{(1)} = R_{(0)}^{(1)}
r^{(0)}$.  On this coarser scale, one then solves the {\em error
equation} $A^{(1)}e^{(1)} = r^{(1)}$, relaxing again $\alpha$ times.
(Note that, as discussed in the appendix, the exact solution to the
error equation, when summed with the current iterative solution $v$,
yields the exact solution to the problem $u$.)  The procedure of
passing up the residual to successively coarser scales continues until
a predetermined coarsest scale $n$ is reached.  At this coarsest scale
$n$, one then transfers the solution vector $v^{(n)}$ down to the next
finer scale $n-1$ using a linear {\em interpolation} operator
$I_{(n)}^{(n-1)}$.  The error equation is then relaxed on this scale
$\beta$ times, using $v^{(n-1)} + I_{(n)}^{(n-1)} v^{(n)}$ as the
initial guess.  This procedure is followed down to the finest scale,
relaxing $\beta$ times at each scale $i$ with the initial guess
$v^{(i)} + I_{(i+1)}^{(i)} v^{(i+1)}$, where finally one relaxes
$\beta$ times on the original linear equation to obtain an approximate
solution to the overall problem.

The procedure described above is the {\em V-cycle of the multigrid
algorithm}.  V-cycles may be repeated in succession to produce a
solution of any desired accuracy.

\subsection{Theory for Interpolation and Restriction Operators}

The interpolation and restriction operators should obey two
conditions, known as the {\em variational properties}.  The first
condition, the {\em Galerkin condition}, requires that 
\begin{eqnarray} \label{galer}
A^{(n+1)} = R_{(n)}^{(n+1)} A^{(n)} I_{(n+1)}^{(n)}.   
\end{eqnarray}
It is a recipe for the linear operator to be used in relaxations on
the next coarser scale.  The second condition states that
\begin{eqnarray} \label{var2}
R_{(n)}^{(n+1)} = c (I_{(n+1)}^{(n)})^T, 
\end{eqnarray}
where $c$ is a scalar constant.  This is the recipe for the
restriction operator, up to an overall scalar constant, given the
interpolation operator.

\section{Interpolet Theory} \label{interp}

In this section, we briefly review interpolet theory based on the
discussion in Lippert, Arias and Edelman [1].

In interpolet theory, functions varying slowly over integer
length scales can be closely approximated as linear combinations of
interpolating functions,
\begin{eqnarray} \label{eqn:fifteen}
f(x) = \sum_n f_n {\cal I}(x-n),
\end{eqnarray}
where the $f_n$ are the expansion coefficients and the ${\cal I}(x-n)$
are functions with compact support, also known as {\em interpolets}.

Interpolets are a basis set combining wavelet theory with finite
element analysis.  In place of the orthonormality condition common in
wavelet theory, interpolets have cardinality and interpolation from
finite element analysis as their characteristic properties.  We will
first discuss the properties which interpolets share with finite
elements and then we will describe those properties which they share
with traditional wavelets.

Cardinality means that the values of a function are zero at all
integers except for zero,
\begin{eqnarray} \label{eqn:sixteen}
{\cal I}(n) = \delta_{n0}, \, \, \, \mbox{for all integers $n$}.
\end{eqnarray}
As a consequence of this condition, the function formed by the linear
expansion in Eq.  (\ref{eqn:fifteen}) will match exactly the value of
the original continuous function at the integer grid points when the
expansion coefficients are taken to be the values of $f(x)$ at the
integers, $f_n \equiv f(n)$,
\begin{eqnarray} \label{eqn:seventeen}
f(n) = \sum_m f_m {\cal I}(n-m) = \sum_m f_m \delta_{nm} = f_n.
\end{eqnarray}
Interpolation is the further condition that Eq.  (\ref{eqn:fifteen})
reproduces any polynomial, up to order L for all $x$,
\begin{eqnarray} \label{eqn:eighteen}
x^k = \sum_n n^k {\cal I}_L(x-k), \, \, \, k=0, 1,..., L,
\end{eqnarray}
where subscript $L$ denotes the order of interpolation.  Given the
values of a function at the integers, interpolets then provide a
compact estimate correct to $L^{th}$ order of the values of the
function at the half integers through
\begin{eqnarray} \label{eqn:nineteen}
f(\frac{x}{2}) = \sum_m f_m {\cal I}_L(\frac{x}{2}-m).  
\end{eqnarray}
The matrix representations for the interpolation operators defined by
this relation for both first order interpolets ($L = 1$) and third
order interpolets ($L = 3$) are given below.  Note that the $L=1$ case
gives the familiar linear interpolation procedure.  and that the $L=3$
case is distinct from the interpolation given by traditional cubic
$B-$splines.  Interpolet theory thereby provides guidance in selecting
natural interpolation operators for use in the multigrid algorithm.

$$
{\cal I}_{L=1} = \left(
\begin{array}{c c c c c c}
\ddots & & & & & \\ 
 & \frac{1}{2} & & & & \\
 & 1 & & & & \\
 & \frac{1}{2} & \frac{1}{2} & & & \\
 & & 1 & & & \\
 & & \frac{1}{2} & \frac{1}{2} & & \\
 & & & 1 & & \\
 & & & \frac{1}{2} & \frac{1}{2} & \\
 & & & & 1 & \\
 & & & & \frac{1}{2} & \\
 & & & & & \ddots \\ 
\end{array} \right),
$$
$$
{\cal I}_{L=3} = \left( 
\begin{array}{c c c c c c}
\ddots & & & & & \\
 & -\frac{1}{16} & & & & \\
 & 0 & & & & \\
 & \frac{9}{16} & -\frac{1}{16} & & & \\
 & 1 & 0 & & & \\
 & \frac{9}{16} & \frac{9}{16} & -\frac{1}{16} & & \\
 & 0 & 1 & 0 & & \\
 & -\frac{1}{16} & \frac{9}{16} & \frac{9}{16} & -\frac{1}{16} & \\
 & & 0 & 1 & 0 & \\
 & & -\frac{1}{16} & \frac{9}{16} & \frac{9}{16} & \\
 & & & 0 & 1 & \\
 & & & -\frac{1}{16} & \frac{9}{16} & \\
 & & & & 0 & \\ 
 & & & & -\frac{1}{16} & \\
 & & & & & \\
 & & & & & \ddots \\
\end{array} \right)
$$

In addition to properties from finite element analysis, interpolets
satisfy the central condition of wavelet theory, the {\em two-scale
relation}.  This relation states that every coarse scale interpolet is
expressible as a linear combination of finer scale interpolets.
Interpolets therefore interpolate themselves {\em exactly},
\begin{eqnarray} \label{eqn:twenty}
{\cal I}_L(x) = \sum_n c_n {\cal I}_L(2x-n).
\end{eqnarray}
Here, the ${\cal I}_L(x)$ is the coarse scale interpolet, the ${\cal
I}_L(2x-n)$ are fine scale interpolets, and the $c_n$ are the
expansion coefficients.  (Note that from the cardinality property, we
have that the $c_n$ are just the values of the interpolet at the
half-integers $c_n={\cal I}_L(n/2)$.)  The significance of this
condition is to ensure that a basis made from interpolets of varying
scales always provides a very uniform description of space.  This
point is described in detail the next section.

\renewcommand{\thetable}{\Roman{table}}
\begin{table}\begin{center}
\begin{tabular}{|c||c|c|c|c|c|c|c|c|c|} \hline
x & -2 &-$\frac{3}{2}$ & -1 & -$\frac{1}{2}$ & 0& $\frac{1}{2}$ & 1 & $\frac{3}{2}$ & 2  \\ \hline \hline
l=1 & & & 0 & $\frac{1}{2}$ & 1 & $\frac{1}{2}$ & 0 & & \\ \hline
l=3 & 0 & -$\frac{1}{16}$ & 0 & $\frac{9}{16}$ & 1 &$\frac{9}{16}$ & 0 & -$\frac{1}{16}$ & 0 \\ \hline
\end{tabular}
\end{center}
\caption{Values of interpolets of orders $l=1,3$ at various
points $x$ within their compact support.  Values not shown are zero.}
\label{hgtableI}
\end{table}

The self-interpolation property may be applied recursively to express
an interpolet in terms of interpolets on arbitrary finer scales,
\begin{eqnarray} \label{eqn:twentyandhalf}
{\cal I}_L(x) & = & \sum_n \sum_m \, ...  \, \sum_k c_n c_m \, ...  \,
c_k \\
&& \times \ \ {\cal I}_L(2^a x - 2^{a-1} n -2^{a-2} m \, ...  \, - k)
\nonumber \\
& = & \sum_k {\cal I}_L(k/2^a) {\cal I}_L(2^ax-k).  \nonumber
\end{eqnarray}
where $a$ is a positive integer that determines the fine scale on
which the original interpolet is expanded.  This relation gives a
procedure for determining the values of any interpolet recursively
once given the values of the $c_n$.  Table \ref{hgtableI} presents these
expansion coefficients from which the interpolets may be constructed
through the two-scale relation.

\section{Non-orthogonal Multiresolution Analysis} \label{mra}

\begin{figure} 
\begin{center}
\scalebox{0.30}{\includegraphics{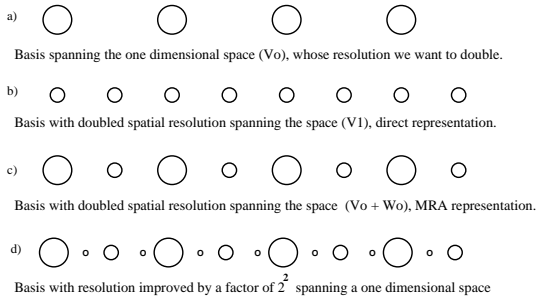}}
\end{center}
\caption{This figure illustrates pictorially how an MRA basis is created.  The bigger circles represent the coarse scale interpolets, while the smaller circles are for the fine scale interpolets.}
\label{fig:explainmra}
\end{figure}

In this section, we present a brief discussion of multiresolution
analysis (MRA) of interpolet and wavelet theories [1],[3], and the
application of MRA to non-uniform grids such as the ones employed in
electronic structure problems.

Assume that there is a basis spanning a space $V_0$ as in Figure
\ref{fig:explainmra}a.   We refer to this space $V_0$ as the {\em
coarse real space} and this representation as the {\em direct
representation}.  In this space, any function $f(x)$, varying slowly
over the given grid spacing, can be approximated with a linear
combination of interpolets of the appropriate scale as explained in
the previous section,
\begin{eqnarray} \label{eqn:twentythree}
f(x) = \sum_n f_n {\cal I}_L(x-n).
\end{eqnarray}
To double the resolution of this basis, we can double the number of
grid points and decrease the scale of the basis functions by a factor
of two, as Figure \ref{fig:explainmra}b illustrates.  In this new
space, $V_1$, which we refer to as the {\em fine real space}, any
function $f(x)$ varying slowly over the new grid spacing can be
approximated with a linear combination of fine scale interpolets with
half the compact support as
\begin{eqnarray} \label{eqn:twentyfour}
f(x) = \sum_n \tilde f_n {\cal I}_L(2x-n).
\end{eqnarray}

General wavelet theory provides another tool to enhance the spatial
resolution of the basis in Figure \ref{fig:explainmra}a, {\em
locally}.  This tool is known as {\em multiresolution analysis} (MRA).
MRA states that to double the resolution of a given basis, we may add
to the existing basis a new set of functions spanning a space $W_0$,
such that $V_0 + W_0 = V_1$.  For this condition to hold, we need to
prove the two constraints that these spaces should satisfy, namely
$V_0 + W_0 \supset V_1$ and $V_0 + W_0 \subset V_1$.  In order for the
first condition to hold, it is sufficient that the interpolets satisfy
the {\em two-scale relation}, Eq.  (\ref{eqn:twenty}) of Section
\ref{interp}.  In order to prove the second condition, one must show
that any interpolet ${\cal I}_L(2x-n) \in V_1$ can be written as a
linear combination of interpolets in the space $V_0 + W_0$.  (This is
proven for interpolets in [1].)  The new basis spanning the space $V_0
+ W_0$ is illustrated in Figure \ref{fig:explainmra}c.  Henceforth, we
will refer to such a mixed basis space as the MRA space and such a
representation as the {\em MRA representation}.

The MRA basis can be extended to include not only the next finer scale
but other scales of increasing fineness, as Figure
\ref{fig:explainmra}d illustrates.  Adding $W_i$'s ($i = 0, 1, 2,...,
N$) to $V_0$ improves the resolution of the original space by a factor
of $2^{N+1}$, $$V_{N} = V_0 + W_0 + W_1 + ...  + W_i + ...  + W_{N-1}.$$

Multiresolution analysis is especially useful when the problem
requires only local regions of high resolution and only a small
fraction of the expansion coefficients for the MRA basis are
significant.  Under these circumstances, one needs to employ only a
small subset of the full MRA basis.  We shall refer to the points in
space associated with the remaining basis functions as the {\em
non-uniform grid} for the problem.  In electronic structure problems,
for instance, only a small spherical region around the atomic core
requires high resolution to describe the most rapid oscillations in
the electronic wave functions.  Such problems are best addressed using
non-uniform grids where high densities of grid points are concentrated
in concentric spheres with differing resolution surrounding the atomic
nuclei.  The advantage of using such an MRA basis appears when the
nuclei move.  New grids do not have to be generated, and associated
Pulley forces calculated, as the nuclei move.  Rather, high resolution
MRA basis functions may be simply turned on and then off as nuclei
pass by.  [1] provides a more detailed explanation of this
application.

\section{Combining the Multigrid Algorithm with Interpolets and MRA} 
\label{img}

We have so far described the multigrid algorithm and the interpolet
basis.  In this section, we discuss the rationale for combining the
multigrid algorithm with interpolets.  The result of Section
\ref{imgss} will be an algorithm in the direct representation.  We
will discuss multiresolution algorithms in Section \ref{dumb}.

\subsection{Choice of Interpolation and Restriction Operators} \label{imgss}

Interpolets provide a natural prescription for an interpolation
operator to be used in the multigrid algorithm.  However, multigrid
also requires an appropriate restriction operator so that both
variational properties are satisfied.  In this section we will see
that the appropriate restriction operator is just the transpose of the
interpolet interpolation operator.  This comes about because of the
form of linear operators generated by applying the Galerkin technique
to the interpolet basis.

Let the following be the $d$-dimensional linear equation to be solved,
\begin{eqnarray} \label{eqn:mrmgone}
\hat {\cal O} \phi(x) = \rho(x),
\end{eqnarray}
where $\hat {\cal O}$ is a linear operator and $\phi(x)$ and $\rho(x)$
are functions of the $d$-dimensional variable $x$.  If we are to solve
this equation in an interpolet basis, $\phi(x)$ and $\rho(x)$ are then
expanded as,
\begin{eqnarray} \label{eqn:mrmgtwo}
\phi(x) = \sum_n \tilde \phi_n {\cal I}_L(x-n), \, \, \rho(x) = \sum_n
\tilde \rho_n {\cal I}_L(x-n).
\end{eqnarray}
Substituting these expansions into the linear equation and applying
$\int dx \, \, {\cal I}_L(x-m)$ to both sides yields
\begin{eqnarray} \label{eqn:mrmgfour}
\sum_n A^{(n)}_{mn} \tilde \phi_n = \sum_n U^{(n)}_{mn} \tilde \rho_n,
\end{eqnarray}
where
\begin{equation}
A^{(n)}_{mn} \equiv \int dx \, \, {\cal I}_L(x-m) \, 
\, \hat {\cal O} {\cal I}_L(x-n)
\end{equation}
is the interpolet form for the linear operator on scale $n$, and
\begin{equation}
U^{(n)}_{mn} \equiv \int dx \, \, 
{\cal I}_L(x-m) {\cal I}_L(x-n).
\end{equation}

Using the two scale relation, we may relate this operator to the
linear operator on the next coarser scale $A^{(n+1)}$,
\begin{eqnarray} 
A^{(n+1)}_{mn} & \equiv & \int dx \,\, {\cal I}_L(x/2-m) \hat {\cal O} {\cal
I}_L(x/2-n) \nonumber \\
& = & \sum_{kl} c_k c_l \int dx \, \, {\cal I}_L(x-2m-k)
\, \, \hat {\cal O} {\cal I}_L(x-2n-l) \nonumber \\
& = & \sum_{kl} c_k A_{2m+k, 2n+l}^{(n)} c_l \nonumber \\
A^{(n+1)} & = &  {\cal I}_L^T A^{(n)} {\cal I}_L. \label{eqn:mrmgfive}
\end{eqnarray}
In the final line, we convert the relationship between the operators
on the two scales into a matrix equation.  From
Eq. (\ref{eqn:mrmgfive}), we may construct a set of operators that
automatically satisfy both variational properties, the Galerkin
condition and the full weighting condition.  The appropriate set of
operators are as follows.  First, the linear operators $A^{(n)}$
representing $\hat {\cal O}$ will be those generated from the
interpolets using the standard Galerkin procedure.  Second, the
interpolation operators ${\cal I}_L$ will be those generated from the
interpolets as discussed in Section \ref{interp}.  Finally, this
analysis shows that to complete the set, the restriction operators
must be $R = {\cal I}_L^T$.  

Implementing the interpolet basis in the multigrid algorithm is then
using the linear operators $A^{(n)}$ to solve Eq. (\ref{m1}) and the
error equation, and using the matrices ${\cal I}_L$ and ${\cal I}_L^T$
as the interpolation and the restriction operators, respectively.

\subsection{Synthesis of MRA into the Interpolet Multigrid Algorithm} \label{dumb}

In this section, we discuss how we may generalize the algorithm that
combines the interpolet basis with the multigrid algorithm to
multiresolution bases.  We also discuss briefly how this multigrid
algorithm in the MRA bases may be applied to non-uniform grids in an
efficient manner.

\begin{figure} 
\begin{center}
\scalebox{0.30}{\includegraphics{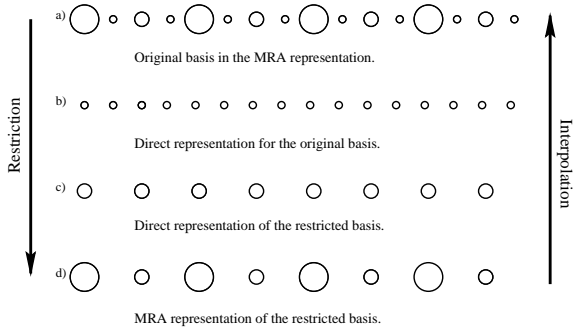}}
\end{center}
\caption{Interpolation and restriction operations carried out in the MRA representation. The arrows indicate the direction to follow to carry out interpolation and restriction. The larger (smaller) circles represent the coarser (finer) scale interpolets.}
\label{usemra}
\end{figure}

We will now discuss two approaches for carrying out interpolation and
restriction for data in the MRA representation illustrated in Figure
\ref{usemra}a.  The first approach is a direct approach and is
equivalent to the method explained in the previous section.  To carry
out restriction in this approach, one first changes from the MRA
representation to the direct representation (the process carrying data
from Figure \ref{usemra}a to Figure \ref{usemra}b.  One then carries
out the restriction operation in the usual way by applying the
previously described linear operator ${\cal I}_L^T$ (making the
transformation Figure \ref{usemra}b$\rightarrow$\ref{usemra}c).
Finally, one converts back from the direct representation to the MRA
representation (\ref{usemra}c$\rightarrow$\ref{usemra}d).
Interpolation within this approach also consists of three
steps. First, we change from the MRA representation to the direct
representation (\ref{usemra}d$\rightarrow$\ref{usemra}c), then we
carry out interpolation with the usual operator ${\cal I}_L$
(\ref{usemra}c$\rightarrow$\ref{usemra}b), and finally we convert back
from the direct representation to the MRA representation
(\ref{usemra}b$\rightarrow$\ref{usemra}a).  We refer to this approach
of changing from the MRA representation to the direct representation
in order to carry out the interpolation and the restriction operations
as the {\em operator method}.

\begin{figure} 
\begin{center}
\scalebox{0.40}{\includegraphics{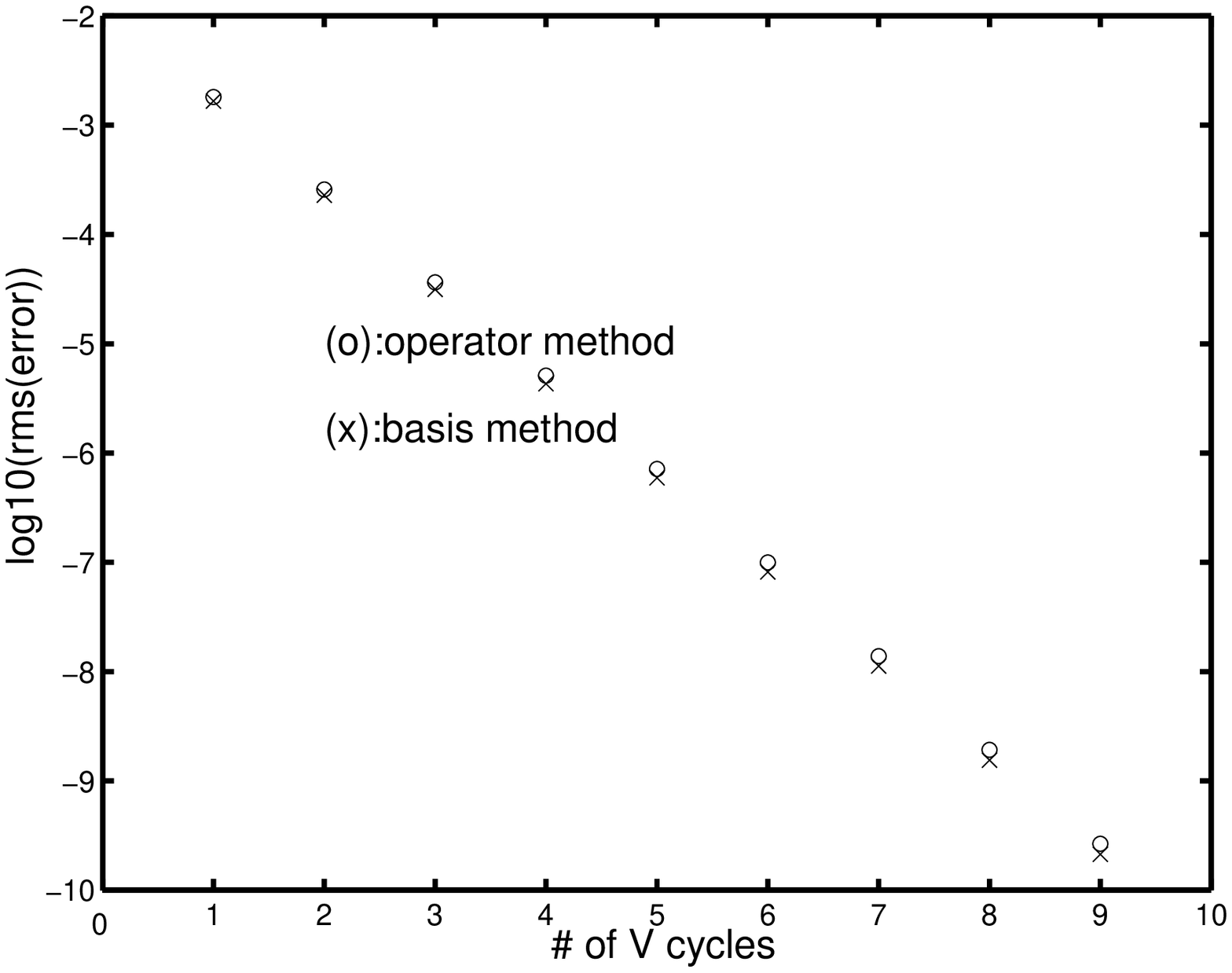}}
\scalebox{0.40}{\includegraphics{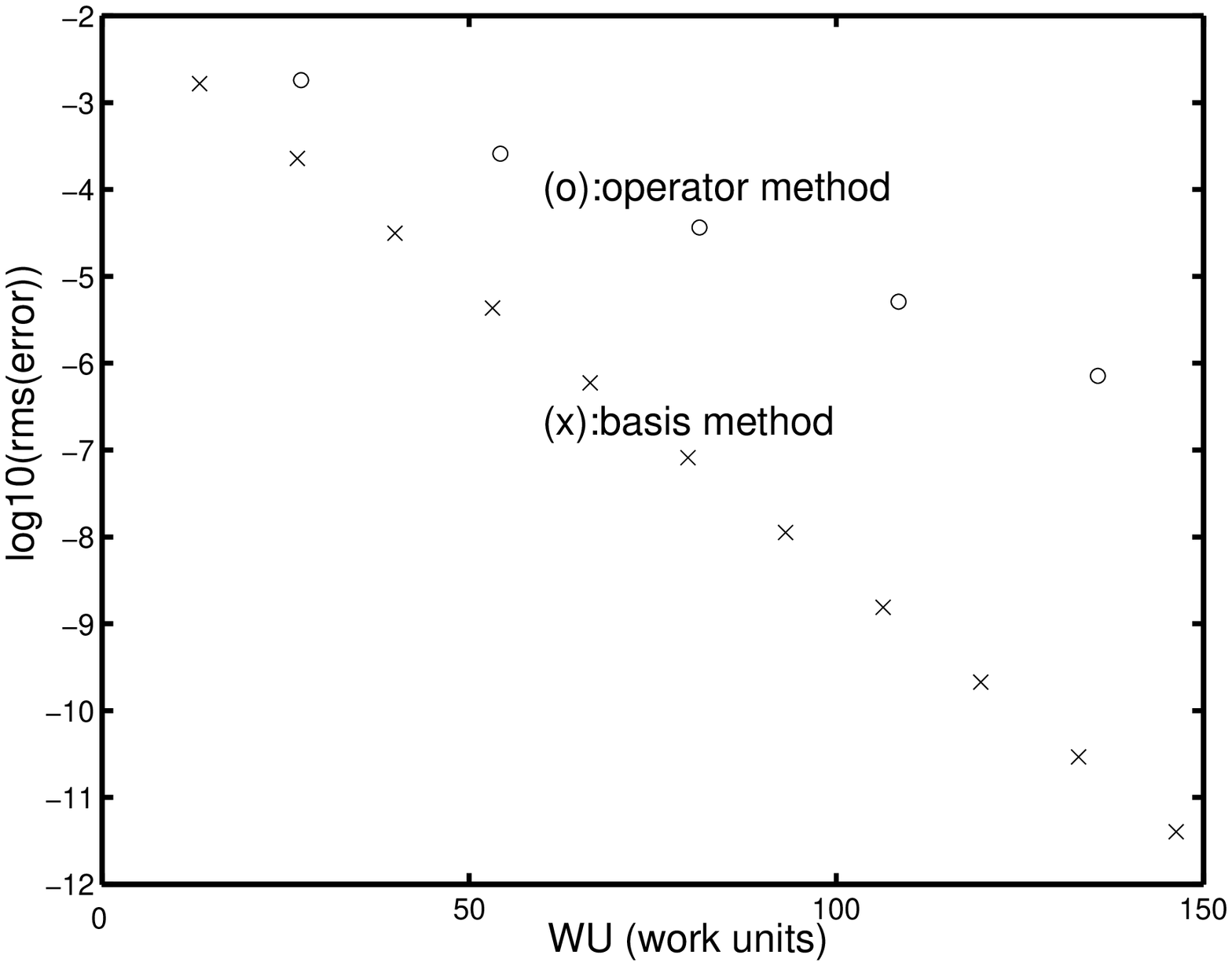}}
\end{center}
\caption{The operator method compared to the basis method.  The first figure represents the convergence rate vs the number of V-cycles and the second figure shows that the convergence rate of the basis method is computationally more efficient when the number of work units are considered.  The above graphs are obtained by solving the Poisson's equation on a grid of 128 points, using first order interpolets.}
\label{dumbo}
\end{figure}

The second approach is an improvement upon the operator method.  This
approach uses the properties of the MRA basis to carry out the
operations of interpolation and restriction entirely within the MRA
representation.  We refer to this second method as the {\em basis
method}.  It uses the general physical meaning of restriction and
interpolation to exploit the interpolet basis.  Restriction is the
elimination of the fine information from a given basis and mapping it
onto a coarser one.  In the interpolet basis, this fine information
corresponding to the high frequency modes is carried by the finest
interpolets.  Therefore, restriction in this basis corresponds to the
obviation of all the coefficients for the finest interpolets.
Interpolation maps coarse information onto a finer grid.  In the
interpolet basis, the coarse information is represented by the coarse
interpolets.  Thus, interpolation leaves the coarse interpolets
unaffected, padding the coefficients for all finer interpolets with
zeros.

The operators in this new method satisfy the Galerkin condition since
the interpolation operator in the MRA representation consists of a
block identity matrix and a block zero matrix for the coarse and fine
informations respectively.  Figure \ref{dumbo}a illustrates that the
operator and the basis methods are equivalent with the same
convergence rates per V-cycle.

Note that the basis basis method has three distinct advantages over
the operator method.  First, the basis method requires no
matrix-vector multiplications to implement interpolation and
restriction.  Figure \ref{dumbo}b illustrates that the basis method is
superior to the operator method on a per flop basis.  Second, the
basis method provides the abilities to interpolate and to restrict
over many scales in a single step.  Finally, the basis method is
easily generalized to problems on non-uniform grids.  On non-uniform
grids, restriction is merely obviation of finer scale coefficients and
interpolation is padding finer scale coefficients with zeros.

\section{The Hopgrid Algorithm} \label{hgrid}

In this section we introduce the Hopgrid algorithm, which is the
synthesis of the multigrid algorithm and the interpolet basis in the
MRA representation.  We present Hopgrid as an algorithm to solve
discretized linear differential equations on interpolet bases in MRA
representation.

\begin{figure} 
\begin{center}
\scalebox{0.40}{\includegraphics{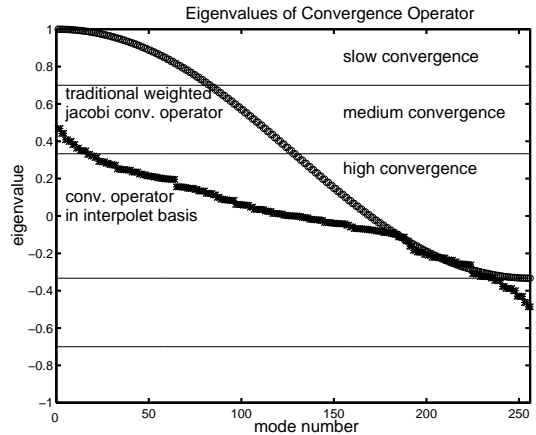}}
\end{center}
\caption{Eigenvalue spectrum of the convergence operators for the weighted Jacobi method in the traditional multigrid basis and the interpolet MRA basis for a problem of 256 grid points.  For the eigenvalue spectrum of the traditional weighted Jacobi operator, w=2/3, and for the weighted Jacobi operator in the interpolet MRA basis, w=1.}
\label{evals}
\end{figure}

The efficiency of using the interpolet MRA basis in the multigrid
algorithm appears when we look at how the error behaves at each step.
The error vector at iteration step $n$ is
\begin{eqnarray} \label{conv1}
e_{(n)} = A ^{-1} f - v_{(n)}.
\end{eqnarray}
Substituting the definition of the residual, $r = f - A v$ and Eq.
(\ref{m2}) into Eq.  (\ref{conv1}) returns the following expression,
which states that error at each step is multiplied by a convergence
operator
\begin{eqnarray} \label{conv2}
e_{(n+1)} = \hat {\cal C} e_{(n)}
\end{eqnarray}
where $\hat {\cal C} \equiv 1 - w D^{-1} A$.  The eigenvalues of this
operator determine how the error behaves at each iterative step.
Figure \ref{evals} illustrates the eigenvalue spectrum of the
convergence operators for both the traditional weighted Jacobi method
and the weighted Jacobi method in the interpolet MRA basis.

The convergence bands shown in Figure \ref{evals} are chosen for
convenience to describe the effects of the eigenvalues on the error
modes.   The {\em high convergence band} is defined to be the region in
which the choice of $w$ for the weighted Jacobi recursion guarantees
that about half of the eigenvalues of the convergence operator of the
traditional weighted Jacobi method lie in the vicinity of zero.   The
errors which are multiplied by these eigenvalues approach zero at a
high rate.   The {\em medium convergence band} and the {\em slow
convergence band} are the regions where the convergence of the error
to zero is slower.   The distinction between these two bands in Figure
\ref{evals} is chosen for purposes of illustration.

First, we will consider the convergence of the traditional weighted
Jacobi method.  The choice of weight $w$ in the method affects the
number of eigenvalues that lie in each of the bands; however, the
operator $D^{-1} A$ always has eigenvalues approaching zero, and so
the convergence operator always includes eigenvalues near unity and
modes which converge very slowly.

On the other hand, most of the eigenvalues of the convergence operator
in the interpolet MRA basis are clustered in the high convergence
band.  The reason for this behavior of the eigenvalues is the fact
that the flow of information on the interpolet MRA basis is stronger
than the flow of information in the basis which the traditional
multigrid algorithm uses.  The superiority in communication in the MRA
basis comes from the fact that the MRA basis includes functions with
support spanning the entire spatial extent of the problem.

The information given in Figure \ref{evals} shows that after one
weighted Jacobi relaxation in the interpolet MRA basis, most of the
error modes are eliminated.  Only the modes that correspond to the
eigenvalues at the edges of the spectrum survive after relaxation.
Numerical experiments show that these surviving error modes have low
spatial frequencies and so may be eliminated very efficiently, using
the basic ideas underlying the traditional multigrid algorithm.

\begin{figure} 
\begin{center}
\scalebox{0.40}{\includegraphics{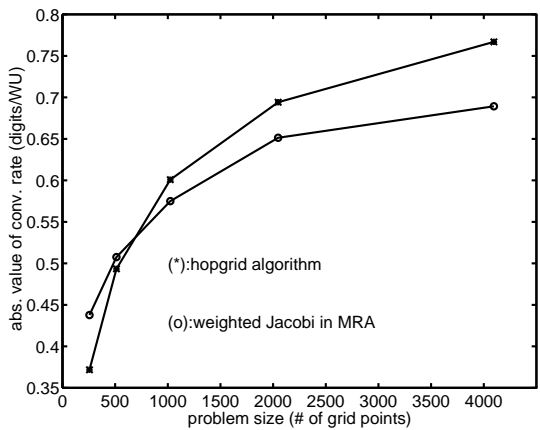}}
\end{center}
\caption{Comparison between the weighted Jacobi method in the MRA representation and the Hopgrid algorithm for the solution of Poisson's equation on grids of several different sizes.}
\label{hgwj}
\end{figure}

We now exploit these observations to produce a new algorithm.  The
basic outline of the algorithm is to first use weighted Jacobi in the
MRA representation to eliminate the error modes that correspond to the
high frequency eigenvalues in the high convergence band, and then to
restrict the residual to a very coarse spatial scale where the error
equation is solved exactly, eliminating the remaining error modes.
The fact that we {\em hop} over many scales in order to produce the
error vector gives this new algorithm its name, the {\em Hopgrid
Algorithm}.  Below we shall refer to the scale on which we solve the
error equation as the {\em hop scale}, $h$.  Figure \ref{hgwj} shows
the comparison between the convergence rates in digits/WU (work unit)
vs problem size for the Hopgrid algorithm and the weighted Jacobi
method in the MRA representation.  As this figure illustrates, the
hopping over many scales to calculate the error improves the
convergence rate, especially with increasing problem size.

The detailed procedure for the Hopgrid algorithm is as follows.
First, one applies weighted Jacobi in MRA representation once on the
finest scale ($\alpha$=1) to find an approximate solution vector
$v^{(0)}$.  Then, to reduce the computational overhead of the
algorithm, instead of calculating all the entries of the residual
$r^{(0)} = f - A v^{(0)}$, we calculate only the $2^{h}$ entries which
correspond to the restricted residual on the hop scale $h$.
Furthermore, because these entries depend only weakly on the high
frequency components of $v^{(0)}$, we compute them only with the
$2^{h+2}$ entries of $v^{(0)}$ corresponding to the hop scale with two
levels of refinement.  After {\em hopping} over many scales, the
number of grid points on $h$ is very small compared to the number of
points on the finest scale, so that, now, solving the error
equation, $A^{(h)} e^{(h)} = r^{(h)}$ exactly becomes a negligible
part of the computation.  Finally, we hop again over many scales and
interpolate the resulting $e^{(h)}$ back to the finest scale, again
just padding with zeros.  The sum of the approximate solution before
hopping $v^{(0)}$ and the interpolated error from scale $n$ gives the
improved solution vector $v = v^{(0)} + I_{(n)}^{(0)} e^{(n)}$.  The
hopping procedure may then be applied recursively to reach any desired
level of accuracy.

Numerical experiments show that the best convergence rates are
obtained when the hop scale $h$ is chosen so that $h \leq
\frac{N}{2}$, where $N$ is defined such that the number of grid
points for the original problem is $2^{N}$.

\section{Results} \label{rest}

In this section we compare the efficiency of the Hopgrid algorithm
with the traditional multigrid algorithm.  We use two different
measures of convergence rate for comparison.  The first measure is the
conventional measure in terms of digits per WU [2].  Here, we define
one work unit to be the number of flops it takes to multiply a vector
by the matrix representing the linear operator in the corresponding
basis.  We will see that in terms of this measure the present
algorithm is signficantly superior.  The MRA representation loses part
of its advantage in problems requiring uniform resolution, as on a
uniform grid, MRA operators are far denser, having fractal dimension
1.5, than their single scale counterparts.  Even in this extreme case,
the Hopgrid algorithm overcomes the appearant disadvantage of the
density of the MRA operators.  To demonstrate this we also compare the
convergence rates of the two algorithms on a per flop basis.  Note
that we envision applying the Hopgrid algorithm to non-uniform
problems expresses in MRA representations and the flop count for the
uniform case represents a worse-case scenario.

\begin{figure} 
\begin{center}
\scalebox{0.40}{\includegraphics{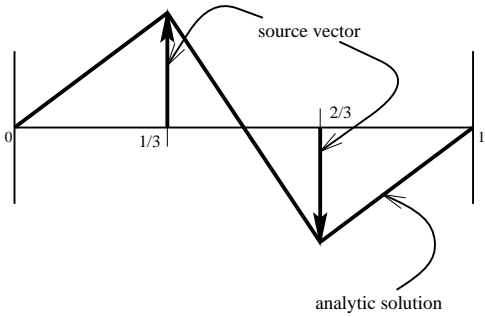}}
\end{center}
\caption{Source vector and the exact solution to Poisson's equation that is used in finding the results shown in this section.}
\label{source}
\end{figure}

The results presented in this section all use third order
interpolets.  The differernatial equation we solve for our tests is the {\em Poisson Equation} in
one dimension on the unit interval with periodic boundary conditions,
\begin{eqnarray} \label{resone}
-\frac{\partial^2}{\partial x^2} \phi(x) = \rho(x).
\end{eqnarray}
As Figure \ref{source} illustrates, we use for testing purposes
two $\delta$-functions of weight one with opposite signs as the source
vector $\rho(x)$.  Given this source
vector, the exact (both analytic and numeric) solution to Poisson's
equation, $\phi(x)$, is the
piece-wise linear function as given in Figure \ref{source}.

Following the procedure given in Section 5.1, we substitute the
interpolet basis expansions for $\phi(x)$ and $\rho(x)$ into
Eq. (\ref{resone}).  In the direct representation, this procedure
returns the matrix equation
\begin{eqnarray} \label{resthree}
\sum_n A_{mn} \tilde \phi_n = \sum_n U_{mn} \tilde \rho_n,
\end{eqnarray}
where $A_{mn} = \int dx \, \, \frac{d}{d x} {\cal I}_L(x-m) \frac{d}{d
x} {\cal I}_L(x-n)$ and $U_{mn} = \int dx \, \, {\cal I}_L(x-m) {\cal
I}_L(x-n)$.  These are the matrix elements employed in the traditional
multigrid calculations, numerical values computed by recursive
application of the two scale relation as described in [1], are
provided in Table \ref{hgtableII}.

\renewcommand{\thetable}{\Roman{table}}
\begin{table}\begin{center}
\begin{tabular}{|c||c|c|c|c|c|c|} \hline
 & & & & & & \\ For $A_{mn}$, values for $\mid m - n \mid$ : & 0 & 1 &
2 & 3 & 4 & 5 \\ & & & & & & \\ \hline \hline & & & & & & \\ First
order interpolet: & -2 & 1 & 0 & & & \\ & & & & & & \\ \hline & & & &
& & \\ Third order interpolet: & $\frac{20}{9}$ & -$\frac{9}{8}$ & 0 &
$\frac{1}{72}$ & 0 & 0 \\ & & & & & & \\ \hline
\end{tabular}\end{center}
\caption{{\em Non-zero matrix elements for the linear operator $A_{mn}$ for the first and the third order interpolets.  The matrix elements for the fifth order interpolets can be calculated in a straight forward manner but are not listed here.}}
\label{hgtableII}
\end{table}

\begin{figure} 
\begin{center}
\scalebox{0.40}{\includegraphics{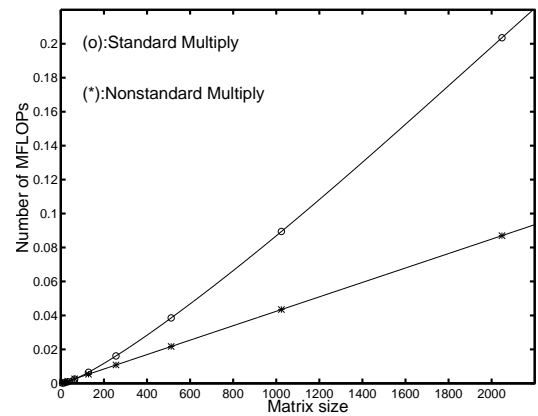}}
\end{center}
\caption{The standard matrix vector multiplication by the linear operator in the MRA basis vs the non-standard multiplication.}
\label{nonst}
\end{figure}

\begin{figure} 
\begin{center}
\scalebox{0.40}{\includegraphics{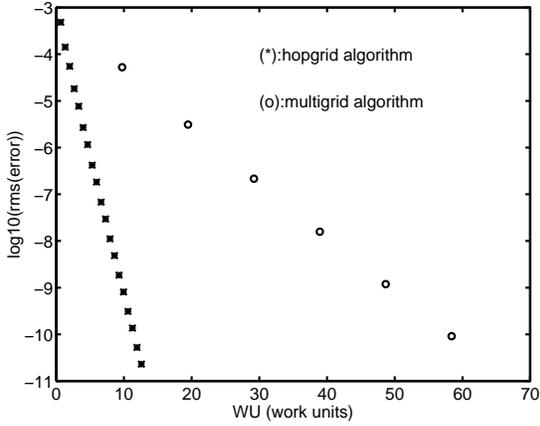}}
\scalebox{0.40}{\includegraphics{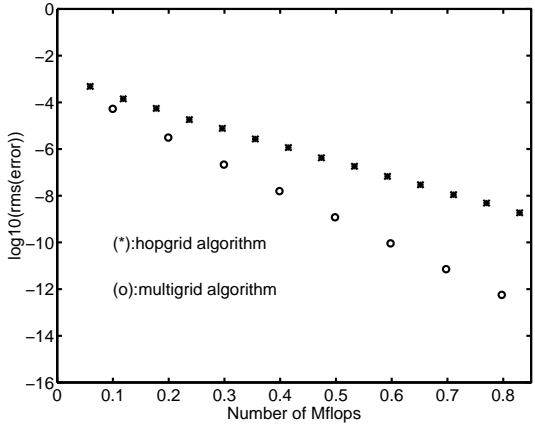}}
\end{center}
\caption{The convergence of the traditional multigrid algorithm versus the convergence of the Hopgrid algorithm for the solution of the Poisson's equation on a grid of 1024 points.}
\label{compare}
\end{figure}

\begin{figure} 
\begin{center}
\scalebox{0.40}{\includegraphics{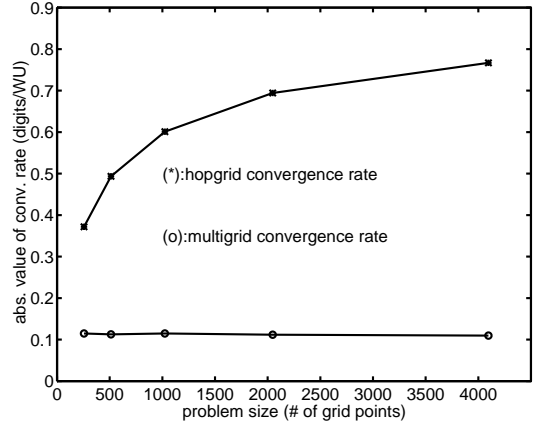}}
\scalebox{0.40}{\includegraphics{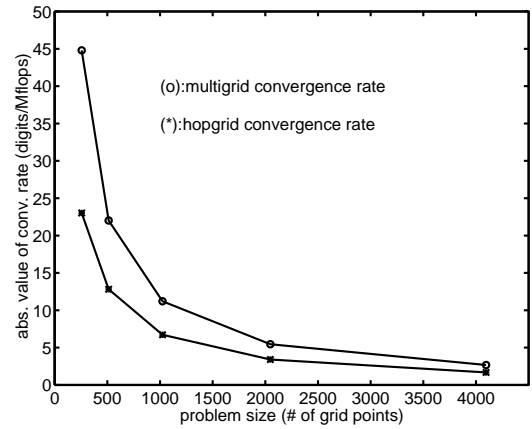}}
\end{center} 
\caption{Comparison between the convergence rates of the traditional multigrid algorithm and the Hopgrid algorithm for increasing problem size.} 
\label{convrate} 
\end{figure}

The Hopgrid calculations require matrix elements of
the linear operator in the MRA represenation.  These elements are
\begin{eqnarray} \label{mramat}
A_{mn}^{mra} = \int dx \, \, \frac{d}{d x} {\cal I}_L(2^{s_m} x-m)
\frac{d}{d x} {\cal I}_L(2^{s_n} x-n),
\end{eqnarray}
where $s_i$ is the scale of basis function $i$.  These values may be
computed from the previous $A_{mn}$ by appropriate change of
integration variable and application of the two scale relation.

Multiplying a vector by this MRA matrix may be performed using one of
two different methods.  The first is straight forward multiplication
by the matrix $A^{mra}$.  Because of the density of the MRA matrix
mentioned above, the number of flops required to apply the operator
within the method grows faster than linearly with the size of the
problem.  The second approach uses the two-scale relation to reduce
the number of operations so that the effort in the matrix
multiplication scales only linearly with the size of the problem [1].
Figure \ref{nonst} compares the number of $Mflops$ needed by the
standard and the non-standard multiplications as a function of problem
size.  We use this superior approach in all of our comparisons.

Figure \ref{compare} shows how the convergence of the Hopgrid
algorithm to the exact solution compares with the convergence of the
traditional multigrid algorithm for a problem of 1024 grid points.  In
Figure \ref{compare}a, the convergence to the exact solution is
plotted against {\em work units} (WU).  Using this comparison the
Hopgrid proceesure is superior to the traditional multigrid algorithm.
When the convergence rates of these two algorithms are compared in
terms of $digits$ per Mflops, as in Figure \ref{compare}b, we find
that despite the increased density of the MRA matrix (by about a
factor of ten), the two methods give comparable results on a uniform
grid.

Next we investigate the dependence of these convergence rates on
problem size.  In terms of $digits$ per $WU$, Figure
\ref{convrate}a shows that although the convergence rate of the
traditional multigrid algorithm remains nearly constant, the
convergence rate of the Hopgrid algorithm {\em increases} with
increasing problem size.  Figure \ref{convrate}b shows that even for
the uniform case, the efficiency of the Hopgrid algorithm overcomes
the increased complexity of the MRA matrix and results in nearly the
same convergence rate as the traditional method in terms of a direct
floating point operations comparison.  It is in the case of
non-uniform grids, where the application of the MRA matrix requires
far fewer operations, where we expect the maximum from the new
algorithm,

\section{Conclusion}

In this research, our aim was to develop a new algorithm that benefits
from the best combination of a variety of approaches, in particular
the multigrid algorithm, multiresolution analysis from wavelet theory,
and finite element analysis.  The application for the new algorithm
which we have in mind is the numerical solution of problems involving
long-range interactions where an underlying interpolet basis with its
multiresolution properties is needed.  We have shown that the
interpolet theory, based upon multiresolution analysis from wavelet
theory and interpolation properties from finite element theory,
provides a natural and unique choice of interpolation and restriction
operators and an underlying grid for the multigrid algorithm.  Then we
showed how the operations of restriction and interpolation may be
carried out efficiently, without the application of linear operators,
by exploiting the properties of an MRA basis.  Finally, we introduced
the Hopgrid algorithm, which draws upon these results to produce a
multigrid-like method to solve discretized linear equations in
problems expressed in multiresolution wavelet bases.  We have seen
that the convergence rate of the new algorithm in terms of work units
is superior to that of the traditional multigrid apprach and that,
even in the worse-case scenario of a uniform problem, the new
algorithm performs as well as the traditional approach on a direct
floating point operation basis despite the increase complexity of MRA
matrices.

\section{Appendix: Description of the Multigrid Algorithm} \label{appenmg}

In this section, we give a detailed discussion of the traditional
multigrid algorithm. This introduction to multigrid is a review of the
technique as given in {\em A Multigrid Tutorial} by William Briggs
[2].

\subsection{Basic Iterative Methods And the Coarse Grid Correction Scheme}

We first present the weighted Jacobi basic iterative method and the
coarse grid correction scheme which are the core ideas in the
multigrid algorithm.

Assume that the discretized differential equation we wish to solve may
be represented in matrix form as
\begin{eqnarray} \label{eqn:one}
Au = f,
\end{eqnarray}where $A$ is the linear
operator, $u$ is the solution vector, and $f$ is the source vector.
Given an approximation to the solution, which we will refer to as $v$,
the error can be expressed as $e=u-v$. Another measure of error is the
{\em residual} $r$,
\begin{eqnarray} \label{eqn:two}
r \equiv f-Av.
\end{eqnarray}With this definition, 
\begin{eqnarray} \label{eqn:three}
Ae = r
\end{eqnarray}is equivalent to Eq. (\ref{eqn:one}). Henceforth, we shall 
refer to Eq. (\ref{eqn:three}) as the {\em error equation}.

Decomposing the initial matrix $A$ as $A=D-L-U$, where $D$, $-L$, $-U$
are the diagonal, lower triangular and upper triangular elements of
$A$ respectively, Eq. (\ref{eqn:one}) becomes
\begin{eqnarray} \label{eqn:four}
u = D^{-1}(L+U)u + D^{-1}f.
\end{eqnarray}This equation may be solved iteratively with the following 
recursion,
\begin{eqnarray} \label{eqn:five}
v_{(n+1)} = D^{-1}(L+U)v_{(n)} + D^{-1}f.
\end{eqnarray}This recursive algorithm can be improved by choosing an 
appropriate weight $w$, and taking a weighted average of the initial
guess, $v_{(n)}$, and the solution from Eq. (\ref{eqn:five}),
\begin{eqnarray} \label{eqn:six}
v_{(n+1)} = [(1-w)I+wD^{-1}(L+U)]v_{(n)} + wD^{-1}f.
\end{eqnarray}Solving Eq. (\ref{eqn:two}) for $f$ and substituting into 
Eq. (\ref{eqn:six}) gives a simplified form for this recursion, in
terms of the residual alone,
\begin{eqnarray} \label{eqn:seven}
v_{(n+1)} = v_{(n)} + wD^{-1}r_{(n)},
\end{eqnarray}where $r_{(n)}$ is the residual at step $n$. This simplified 
recursion is the core of the basic iterative method known as the {\em
weighted Jacobi method}. The application of this method for solving a
problem is referred to as {\em relaxation}.

\begin{figure} 
\begin{center}
\scalebox{0.40}{\includegraphics{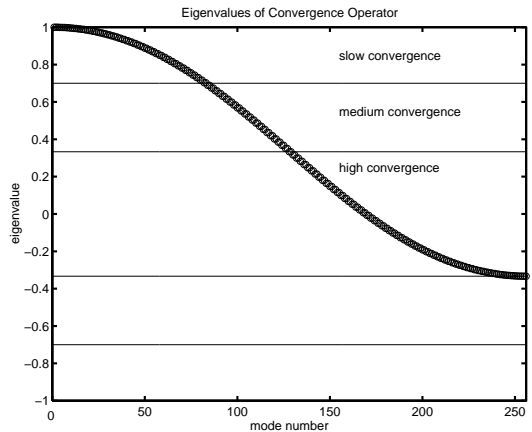}}
\end{center}
\caption{Convergence operator for the weighted Jacobi method.}
\label{wjaceval}
\end{figure}

The weighted Jacobi method and similar basic iterative methods rapidly
eliminate high frequency errors while leaving the smooth components of
the error mostly unaffected. This is called the {\em smoothing
property}.  The reason for this behavior can be explained when we look
at the eigenvalues of the convergence operator $\hat {\cal C} \equiv 1
- w D^{-1} A$. The eigenvalues of this operator determine how fast the
error converges to zero. Fig. \ref{wjaceval} shows that the
eigenvalues of the convergence operator for the weighted jacobi method
lie in various ranges.  The bands are chosen for illustration purposes
and an explanation for these choices are given in the main body of the
paper.

The errors that are multiplied by the eigenvalues in the slow
convergence band correspond to eigenvectors varying slowly in space.
Because the convergence operator $\hat {\cal C}$ does not eliminate
efficiently most of these low frequency errors, this basic iterative
method approach the exact solution very slowly.  A more efficient
method would eliminate all the error modes at equal rates. The
multigrid algorithm has this feature and produces a solution that
approaches the analytic solution with a higher convergence rate.

\begin{figure} 
\begin{center}
\scalebox{0.40}{\includegraphics{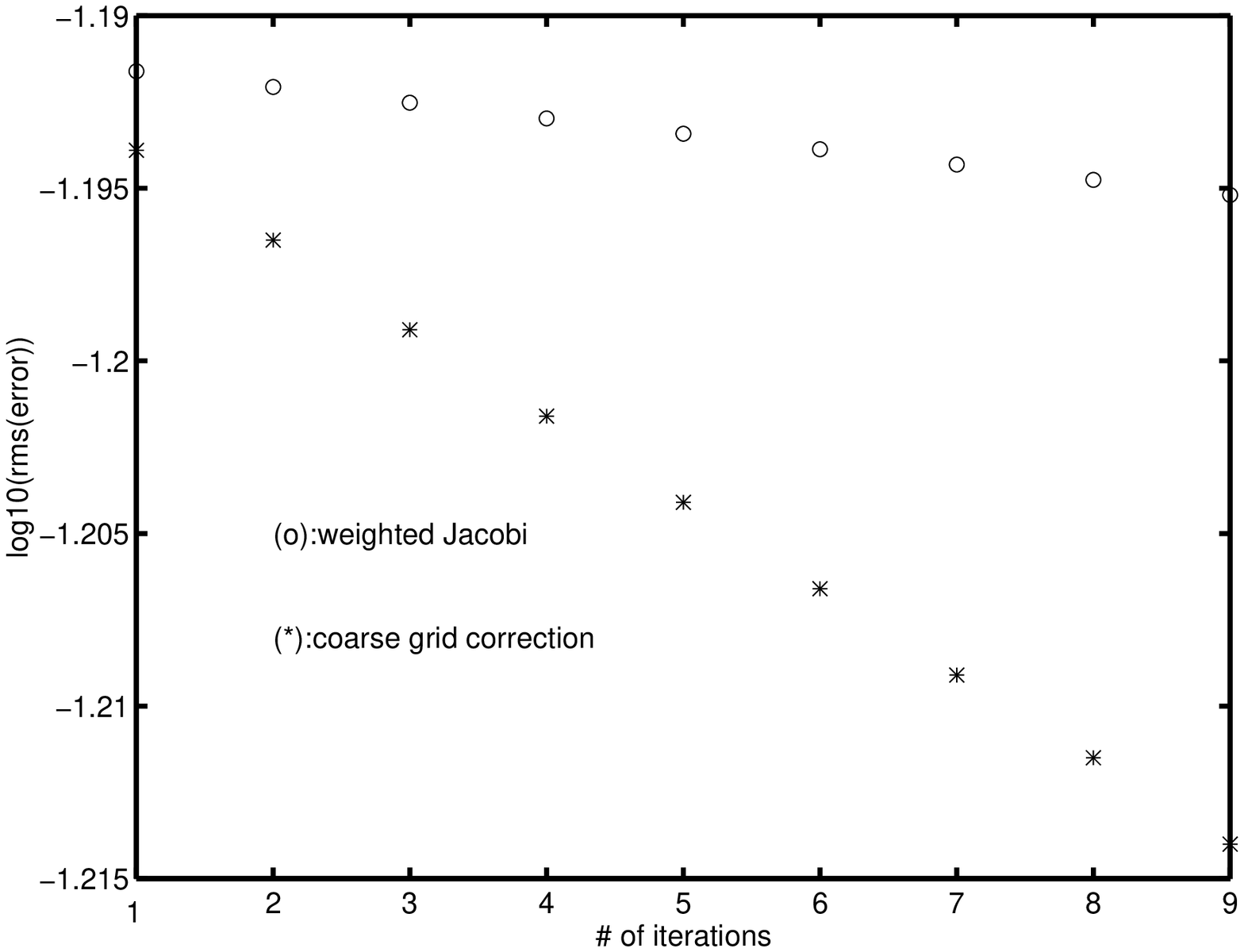}}
\scalebox{0.40}{\includegraphics{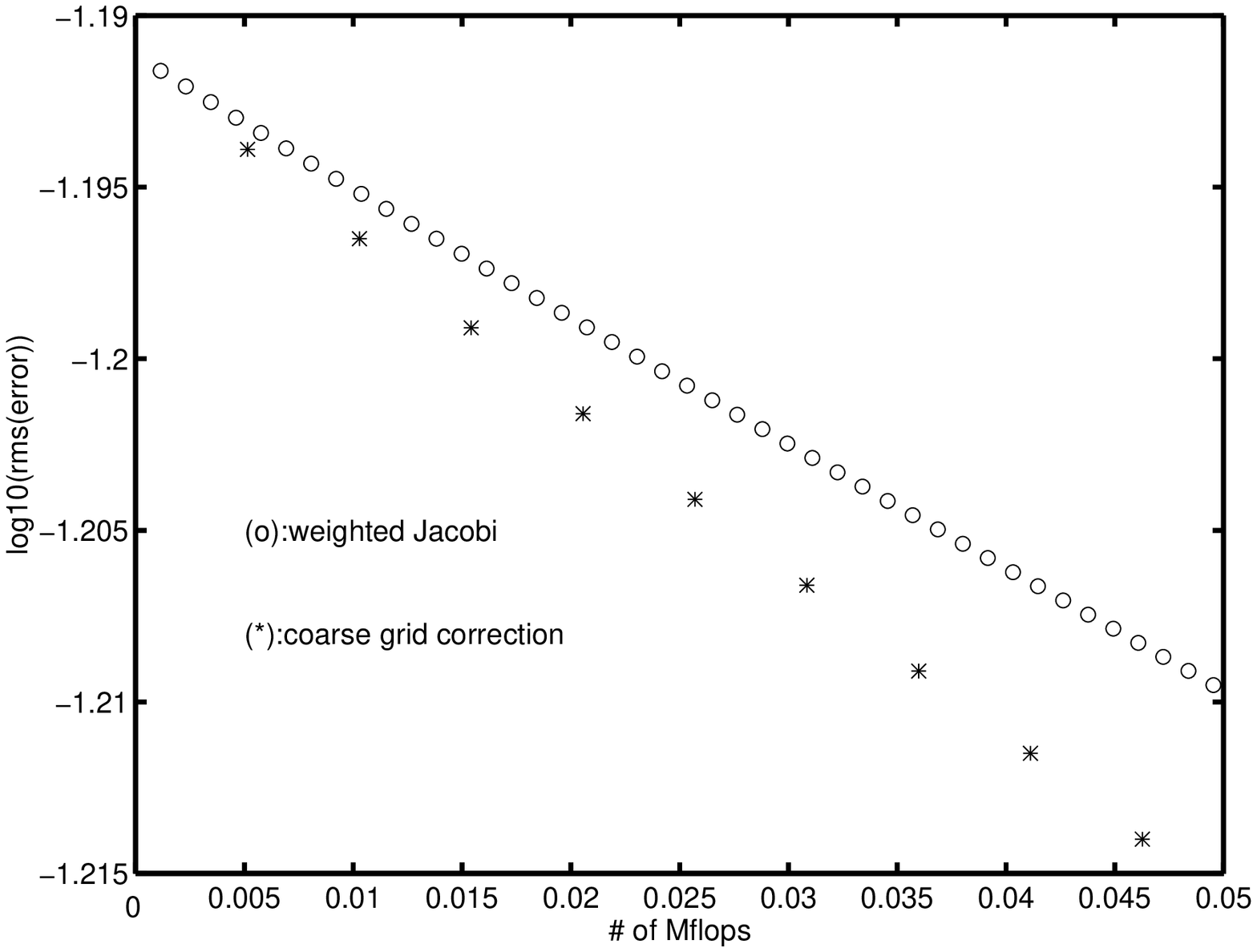}}
\end{center}
\caption{This graph shows the difference between the efficiencies of a pure basic iterative method (weighted jacobi) and the coarse grid correction in finding the solution for the Poisson's equation on a grid of 128 points.}
\label{fig:one}
\end{figure}

The transition from the weighted Jacobi method to the multigrid
algorithm is made by improving upon the basic Jacobi method.  One way
to improve the efficiency of weighted Jacobi relaxations is to start
with a better initial guess. Conceptually, one can do this by solving
the problem on a spatially coarser scale and using this solution as
the initial guess for the original problem.  This procedure is
referred to as {\em coarse grid correction} (CGC).  On the coarse
scale, there are fewer grid points, and on this grid the smooth error
modes from the fine scale now appear higher frequency modes.
Replacing the original problem with a smaller one of mostly high
frequency errors allows the application of the weighted Jacobi method
to solve the problem more efficiently.  This way, the convergence rate
for the total error is accelerated since both the high and the low
frequency error modes are eliminated on the fine and the coarse
scales.

Figure \ref{fig:one} illustrates the accelerated convergence rate of
coarse grid correction scheme compared to the weighted Jacobi method.

\subsection{Full V-cycle Multigrid Algorithm}

The multigrid algorithm simply extends the idea of CGC over a sequence
of scales.  Throughout this section, superscripts are used to refer to
the scale on which that particular level of the problem is solved.

In order to find an approximation to the solution of the linear
equation $Au=f$, we first apply the weighted Jacobi method as in Eq.
\ref{eqn:seven} $\alpha$ times (do $\alpha$ relaxations) on the finest
scale, to obtain an approximate solution $v^{(0)}$ and a residual
$r^{(0)}$.  The residual calculated on this scale is then transferred
to the next coarser scale using a linear operator $R_{(0)}^{(1)}$,
known as the {\em restriction operator}.  In general,
$R_{(i)}^{(i+1)}$ takes a vector from a fine scale $i$, and transfers
it to the next coarse scale $i+1$. (This operator is discussed later
in the appendix.) For the restriction step, we assume that the
relaxation on the finest scale has eliminated most of the high
frequency error components, which lie in the null-space of
restriction, so that little information is lost in the restriction
process. The restricted residual is now
\begin{eqnarray} \label{eqn:eight}
r^{(1)} = R_{(0)}^{(1)} r^{(0)}.
\end{eqnarray}Next we solve the residual equation 
\begin{eqnarray} \label{eqn:nine}
A^{(1)}e^{(1)} = r^{(1)},
\end{eqnarray}where $A^{(1)}$ is the linear operator on the coarse scale, 
performing $\alpha$ relaxations, using $e^{(1)} = 0$ as an initial
guess.  This relaxation produces an approximation to the error on the
fine scale. We could add this error to $v^{(0)}$ to obtain an
approximate solution for the fine scale problem, since $u = v + e$, as
previously explained.  However, the multigrid algorithm continues to
go down to coarser scales and uses CGC to find solutions for the error
equations at each coarser scale to improve the convergence rate.
Following this prescription, we compute the residual on scale $1$ and
transfer it to the next coarser scale, using the restriction procedure
described.  Repetition of such relaxation and restriction steps
continues until the predetermined coarsest scale $n$ is reached.
During each step leading up to the coarsest scale, the generated
approximate solution vectors, the $v^{(i)}$, and the residuals, the
$r^{(i)}$, are stored to be used later in the algorithm. Although the
notation used for the result of the relaxations at each scale is $v$,
it is important to realize that except for the finest scale on which
the original problem is defined, all the vectors $v^{(i)}$ are
approximations to the error since the error equation is used during
relaxations.

At the coarsest scale, another linear operator, an interpolation
operator $I_{(n)}^{(n-1)}$ transfers the approximate solution vector
$v^{(n)}$ down to the next finer scale. In general, the interpolation
operator $I_{(i+1)}^{(i)}$, takes a vector from a coarse scale $i+1$,
and transfers it to the next fine scale $i$. (This operator is also
discussed later.) The interpolated approximate solution vector,
$I_{(n)}^{(n-1)} v^{(n)}$, is the approximation to the error for scale
$n-1$. The sum of the approximate solution $v^{(n-1)}$ and
$I_{(n)}^{(n-1)} v^{(n)}$, give the approximate solution on scale
$n-1$.  However, the interpolation process may introduce high
frequency errors into this solution. We eliminate these errors by
relaxing $\beta$ times on the residual equation using the initial
guess $v_{*}^{(n-1)} = v^{(n-1)} + I_{(n)}^{(n-1)} v^{(n)}$ and the
residual $r^{(n-1)}$ stored from the previous restriction steps as the
source vector. This relaxation produces the final $v^{(n-1)}$ which is
then interpolated to the next finer scale. This procedure is followed
down to the finest scale. On the finest scale, we relax $\beta$ times
on the original equation, Eq.  (\ref{eqn:one}), using the initial
guess $v_{*}^{(0)} = v^{(0)} + I_{(1)}^{(0)} v^{(1)}$ and $f$ as the
source vector. The solution obtained at this final step is the result
of one iteration of the {\em V-cycle of multigrid algorithm} and gives
an approximate solution to the problem.

It is a common procedure to carry out only one relaxation per each
scale. All the results presented here are done with $\alpha = \beta =
1$, and $w = 2/3$.

\subsection{Interpolation and Restriction Operators}

The interpolation and the restriction operators in the multigrid
algorithm obey two constraints, known as the {\em variational
properties}.  The first of these constraints is the {\em Galerkin
condition}, which specifies the form of the linear operator on a
coarser scale.  Assume that the error vector on scale $n$ is in the
range of interpolation, $e^{(n)} = I_{(n+1)}^{(n)} e^{(n+1)}$, for
some vector $e^{(n+1)}$ on scale $n+1$. The residual equation at scale
$n$, $A^{(n)} e^{(n)} = r^{(n)}$, then reads
\begin{eqnarray} \label{eqn:ten}
A^{(n)} I_{(n+1)}^{(n)} e^{(n+1)} = r^{(n)}.
\end{eqnarray}Restricting both sides, we find
\begin{eqnarray} \label{eqn:eleven}
R_{(n)}^{(n+1)} A^{(n)} I_{(n+1)}^{(n)} e^{(n+1)} = R_{(n)}^{(n+1)} r^{(n)},
\end{eqnarray}or
\begin{eqnarray} \label{eqn:twelve}
A^{(n+1)} e^{(n+1)} = r^{(n+1)}
\end{eqnarray}where 
\begin{eqnarray} \label{eqn:thirteen}
A^{(n+1)} \equiv R_{(n)}^{(n+1)} A^{(n)} I_{(n+1)}^{(n)}.
\end{eqnarray}This is precisely the Galerkin condition for the linear 
operator on the next coarse scale. By solving Eq. (\ref{eqn:twelve})
with $A^{(n+1)}$ as the linear operator and the restricted residual
$r^{(n+1)}$, we determine the error vector on scale $n+1$ which when
interpolated becomes the error vector we seek, $e^{(n)}$.

The second variational property specifies the restriction operator
upto a scalar constant, given the interpolation operator,
$R_{(n)}^{(n+1)} = c (I_{(n+1)}^{(n)})^T$, where $c$ is a scalar
constant.  This condition gives the restriction operator the full
weighting property. The full weighting restriction involves taking
some weighted average of values with neighboring points, instead of
simply down sampling, a process known as {\em injection}.


\begin{thebibliography}{0}
\bibitem{1} Ross A. Lippert, Tom\'{a}s Arias, and Alan Edelman, {\em Journal of Computational Physics} {\bf 140}, 278 (1998).

\bibitem{2}W.L. Briggs; {\em A Multigrid Tutorial} (Lancaster Press, Lancaster, Penssylvania, 1987) 
\bibitem{3}I. Daubechies; {\em Ten Lectures on Wavelets} (Siam Press, Philadelphia, 1992)  
\bibitem{4}G. Deslauriers, S. Dubuc; {\em Fractals, dimensions non entieres et applications}, (Masson, Paris, 1987).
\bibitem{6}D.L. Donoho; Interpolating Wavelet Transforms, (Stanford Dept of Statistics  Technical Report 408, Nov 1992).
\bibitem{8}M.B. Ruskai {\em et al}; {\em Wavelets and their Applications}, (Jones and Bartlett, Boston, 1992)
\bibitem{9}C.K. Chui; {\em An Introduction to Wavelets} (Academic Press, Boston, 1992)  
\bibitem{10}C.K. Chui, ed; {\em Wavelets: A Tutorial in Theory and Applications} (Academic Press, Boston, 1992)  
\bibitem{12} S.F. McCormick; {\em Multigrid Methods} (Siam Press, Philadelphia, 1987) 
\bibitem{13} R.D. Cook; {\em Concepts and Applications of Finite Element Analysis} (Wiley, New York, 1992)
\end{thebibliography}
\end{document}